\documentclass[11pt,preprint]{aastex}

\def\deg{$^{\circ}$}

\begin{document}
\title{~~\\ ~~\\ Monitoring the Bi-Directional Relativistic Jets of the Radio Galaxy 1946+708}
\shorttitle{Monitoring 1946+708}
\author{G. B. Taylor\altaffilmark{1}, P. Charlot\altaffilmark{2,3}, R. C. Vermeulen\altaffilmark{4}, and N. Pradel\altaffilmark{4,5}}
\email{gbtaylor@unm.edu}
\altaffiltext{1}{Department of Physics and Astronomy, University of New
Mexico, Albuquerque NM, 87131, USA; gbtaylor@unm.edu. Greg Taylor is also an
Adjunct Astronomer at the National Radio Astronomy Observatory.}
\altaffiltext{2}{Universit\'e de Bordeaux, Observatoire Aquitain des Sciences de l'Univers, 2 rue de l'Observatoire, BP 89, 33271 Floirac Cedex, France}
\altaffiltext{3}{CNRS, Laboratoire d'Astrophysique de Bordeaux -- UMR 5804, 2 rue de l'Observatoire, BP 89, 33271 Floirac Cedex, France}
\altaffiltext{4}{Netherlands Foundation for Research in Astronomy (ASTRON), PO Box 2, 7990 AA Dwingeloo, The Netherlands}
\altaffiltext{5}{National Astronomical Observatory of Japan, 2-21-1 Osawa, Mitaka, Tokyo, Japan}
 
\slugcomment{ApJ, in press}

\begin{abstract}

We report on a multi-frequency, multi-epoch campaign of Very Long
Baseline Interferometry observations of the radio galaxy 1946+708
using the VLBA and a Global VLBI array.  From these
high-resolution observations we deduce the kinematic age of the radio
source to be $\sim$4000 years, comparable with the ages of other
Compact Symmetric Objects (CSOs).  Ejections of pairs of jet
components appears to take place on time scales of 10 years and these
components in the jet travel outward at intrinsic velocities 
between 0.6 and
0.9 c. From the constraint that jet components cannot have intrinsic
velocities faster
than light, we derive $H_0$ $>$ 57 km s$^{-1}$ Mpc$^{-1}$ from 
the fastest pair of components launched from the core. 
We provide strong evidence for the ejection of a new pair of
components in $\sim$1997.  From the trajectories of the jet components we
deduce that the jet is most likely to be helically confined, rather
than purely ballistic in nature. 

\end{abstract}

\keywords{cosmology:distance scale --- galaxies: active --- 
galaxies: individual (1946+708) --- galaxies: jets --- 
radio continuum: galaxies }

\section{Introduction}

Compact symmetric objects (CSOs) are a family of extragalactic radio
sources comprising roughly 3\% of flux-limited samples selected at
high frequencies (Taylor, Readhead, \& Pearson 1996; Peck \& Taylor
2001).  Their defining characteristic is the presence of high
luminosity radio components on both sides of a central engine on
sub-kiloparsec scales with little or no extended emission present.
CSOs typically exhibit terminal hotspots which move apart at
subluminal speeds (Owsianik \& Conway 1998; Gugliucci et al. 2005).  
Jet components en route
between the core and the hot spots where they terminate, appear to move
faster, and superluminal speeds in the jets are seen in a few sources
(Taylor, Readhead \& Pearson 2000).  The jets can be similar or much
brighter than the counterjets.  CSOs on the whole exhibit weak radio
variability, have low radio polarization and low core luminosities.
Their hosts are in general elliptical galaxies (Readhead et al.\ 1996)
though there are a few notable exceptions identified with Seyferts and
quasars.

The general picture that has emerged is that CSOs are young radio
galaxies, with ages between 100 and 10000 years.  Since they are
selected on the basis of relatively unbeamed hot-spots and lobes,
their orientation on the sky is random.  Orientation may still affect
the presence of detectable linear polarization (Gugliucci et
al. 2007), or absorption from atomic gas associated with the hypothesized
gas and dust torus that surrounds the central engine and is thought
to be oriented
perpendicular to the jet axis (Peck et al. 2000; Pihlstr\"om et al. 2003).
However, as Doppler boosting effects are less dominant,
we have the added benefit of 
studying the emission from both sides of the nucleus.

CSOs provide a unique
opportunity to determine the Hubble constant, as a direct measurement 
of the distance to an
object can be obtained by observing angular motion in it, if the
intrinsic linear velocity can be ascertained independently. Basically,
one has the time derivative of a standard ruler, with the additional
constraint that no intrinsic motion can be faster than light.  From
five epochs of observations of the CSO 1946+708 at 5 GHz Taylor \&
Vermeulen (1997) placed a lower limit of H$_0$ $>$ 37 km s$^{-1}$ Mpc$^{-1}$.  
We explore what can be learned from continuing this analysis using
more comprehensive observations.
We also use this symmetric system, for which
the geometry of the jets can be precisely determined, 
to probe the details of jet propagation.  

Kinematic ages for the radio emission in CSOs can be obtained by
measuring the separation speed between the hot spots over time and
comparing this to the overall spatial separation (Owsianik \& Conway 1998;
Gugliucci et al. 2005).  At the moment, the CSO age
distribution seems to be disproportionally stacked towards younger
sources.  However, there are a number of selection effects that need
to be addressed before the meager data collected can be properly
analyzed.  Larger sources are over-resolved by VLBI observations so
that measuring the expansion becomes difficult; more slowly expanding
sources require longer time baselines to obtain age estimates. 
In this work we present observations of the CSO 1946+708 
spanning 16 years. It is
of considerable interest to pursue this line of research to address
the evolution of radio galaxies.

We assume $H_0$ = 71 km s$^{-1}$ Mpc$^{-1}$ and the standard cosmology
so that 1\arcsec = 1.835 kpc and an angular velocity of 1 mas/y = 5.98 c 
at the redshift
(0.1008) of the radio galaxy 1946+708 (Peck \& Taylor 2001).

\section{Observations and Data Reduction}

The observations were carried out at 8.4 GHz and 15.4 GHz over the
course of eight epochs, taken on 1995 March 22, 1996 July 07, 1998
February 06, 1999 July 11, 2001 May 17, and at 8.4 GHz only on 2003
Jan 24/2003 Feb 10, 2006 Feb 18, and 2008 Mar 9 (See Table 1).
Observations in 1995, 1998, and 2001 were observed using all ten
elements of the VLBA of the NRAO, and observations from 2003 on were
observed using a global array including all ten elements of the VLBA,
and five elements from the European VLBI Network (EVN) including the
100m telescope at Effelsberg, the Westerbork phased array, Onsala,
Medicina, and in 2003 only, the telescope at Noto.  Due to problems at
the St. Croix and North Liberty stations, epochs in 1996 and 1999 were
performed using nine VLBA antennas each.  Problems in 2003 prevented
Effelsberg from participating.  Both right and left circular
polarizations were recorded for the first 5 epochs, while the last
three where observed only in right circular polarization.  All epochs
employed 2 bit sampling across IF bandwidths of 8 MHz.  The VLBA
correlator produced 16 frequency channels across each IF during every
2 s integration.  We also include in our analysis the 5 GHz
observations acquired between 1992 and 1995 by Taylor \& Vermeulen
(1997), some of which are contemporaneous with the 8.4 and 15 GHz
observations reported here.

Parallactic angle effects resulting from the altitude-azimuth antenna
mounts were removed using the AIPS task CLCOR.  Amplitude calibration
for each antenna was derived from measurements of antenna gain and
system temperatures during each run.  Delays between the stations'
clocks were determined using the AIPS task FRING (Schwab \& Cotton
1983).  Calibration was applied by splitting the multi-source data set
immediately prior to preliminary editing, imaging, deconvolution and
self-calibration in {Difmap} (Shepherd, Pearson, \& Taylor 1994,
1995).  Multiple iterations of phase self-calibration and imaging were
applied to each source before any attempt at amplitude
self-calibration was made.  The preliminary models developed in
{Difmap} were subsequently applied in AIPS to make phase corrections,
to determine the leakage terms between the RCP and LCP feeds and to
correct for residual phase differences between polarizations.  Final
imaging and self-calibration were performed in {Difmap}.

Model fitting of Gaussian components to the self-calibrated visibility
data was also performed in {Difmap}.  The shapes of the components
were fixed after fitting to the 1995 epoch; in the rest of the epochs
each component was allowed only to move and to vary in flux density in
order to fit the independently self-calibrated visibility data.  Each
frequency band (5, 8.4 and 15.4 GHz) was modelfit independently in
order to allow for differences in component shapes between
frequencies.  For our last three epochs from 2003 to 2008, all global
8.4 GHz observations, an additional pair of components had to be fit
on either side of the core (see \S 4.1).
Uncertainties in the sizes and positions for components
were derived from signal-to-noise ratios and component
sizes (Fomalont 1999).  This assumes that parameters are not 
covariant, which should be the case for the strong, well-separated
components that we present in this analysis of 1946+708.

\section{Results}

While one of the first objects to be confirmed as a bona-fide Compact
Symmetric Object, 1946+708 is in a few ways unusual for this class.
The source is remarkably ``rotation'' symmetric (see Figures 1 and 2).
Narrow, well collimated jets emerge from a compact, flat spectrum
core and bend gradually before terminating in well defined 
hot spots. Faint ``lobes'' of emission are evident at 90\deg\ angles 
to the jets.  Distinct components in the jet are seen to move 
out from the core to the hot spots on both the Northern (jet) 
side and the Southern (counter-jet) side.

Our most uniform, sensitive, and highest resolution sequence
of images is that at 8.4 GHz.  The 8.4 GHz images cover the period
from March 1995 to March 2008, and have noise levels ranging
from 163 $\mu$Jy/beam down to 38 $\mu$Jy/beam (Table 1).

No polarized flux has been detected from 1946+708 at either 
8.4 or 15.4 GHz in any epoch.  Typical 2$\sigma$ limits on the
linearly polarized flux density are $<$100 $\mu$Jy at 8.4 GHz and
$<$170 $\mu$Jy at 15.4 GHz.

\section{Discussion}

\subsection{A New Set of Jet Components}

This source is quite steady in the production of strong and 
distinct jet components.  In addition to the previously studied pairs
N2/S2 and N5/S5, we report here on the discovery of a new set of
components which we name N6/S6 (Fig.~2).  Like the previous sets,
these new components appear to have been launched simultaneously as a
pair, and appear to be moving away from the core in opposite directions.

The angular separation between the new pair (N6 and S6) observed in our
2008 epoch is 1.288 mas.  The apparent proper motion between 
N6 and S6 and the core is difficult to determine owing to 
possible blending of these components with the core, but the total
separation can be compared with the 2006 epoch in which it
was 1.045 mas.  The total apparent expansion speed between the pair
over the past two years is thus
0.238 mas/y or 0.72 c. This puts the ejection age for the new pair
at 1997.3.  Indeed evidence of the new component set 
can be seen in the 2003 epoch, and to a lesser extent in 2001.
The sum of the core flux density and the flux density of the 
new components N6 and S6 during this period (Fig.~3) was rising
steadily, and only seems to be declining later on, as the new 
components move well away from the core.

\subsection{Kinematics in the jets of 1946+708}

In Fig.~4 we plot the positions of the 4 strongest jet components (N2,
N5, S5 and S2) derived from modelfitting elliptical gaussians to the
visibility data at 8.4 GHz.  The well-tracked jet components close to
the core, N5 and S5, appear to move out with the highest apparent
velocity (Table 2 and Fig.~5; and see also the Movie\footnote{The
  Movie is available from http://www.phys.unm.edu/$\sim$gbtaylor/}),
with N5 being the fastest at an apparent velocity of 1.088 c, and S5
being about one third that in very nearly the opposite direction.
Note that these are apparent velocities projected in the plane of the
sky, as opposed to intrinsic, three-dimensional velocities.  The
more distant pair of components (N2/S2) appear to be moving at roughly
half the apparent velocity of the inner (N5/S5) pair.  From this fact
one might be tempted to assert that the jet starts out fast and
decelerates with time.  If this was the case, and the deceleration is
uniform, then we would expect to see the N5/S5 pair slowing down, but
this is not observed.  Even after 12 years of monitoring and a
distance covered corresponding to 10\% of the length of the jet, there
is no compelling evidence for slowing of any of the components.  The
$\chi^2$ values indicate reasonably good fits to a straight line for
all components (the $\chi^2$ values may be systematically better in
declination than in right ascension owing to the north-south
orientation of most of the components).  If current velocities remain
unchanged then the N5/S5 pair will catch up with the N2/S2 pair in
$\sim$75 years, close to the time that they enter the hot spots.
Alternatively, it could be that there is a sudden deceleration of the
jet components at a distance from the core between 3.5 and 9 mas (on
the northern side).  This gap is defined by the minimum distance of 9
mas for N2 in 1995 and the maximum distance for N5 observed in 2008.
This is in the region where the jet bends the most ($\sim40^\circ$)
towards the northern hot spot, so a deceleration region should not be
ruled out prematurely.  Considering the newest pair of components,
N6/S6, their total apparent separation speed is 0.72c, more like that
of the N2/S2 pair (0.63 c), than the N5/S5 pair (1.4 c).  Observations
over the next $\sim$5 years should settle the issue of whether
components emerge with their own, intrinsic set of velocities, or if
they partake in some fixed pattern of acceleration and deceleration.
If there is a fixed, repeating pattern, then it might be possible to
relate it to a helical jet model.  In a helical model, the intrinsic
velocity could be constant while the orientation changes in time to
produce the observed variations in the projected apparent velocities.
Such a model naturally reproduces the rotational symmetry of 1946+708
and has been found to explain observations over many years of the
galactic jet in SS\,433 (Roberts et al. 2008, Hjellming
\& Johnston 1981).

The path of the inner jet components of 1946+708 (N5 and S5) both appear
fairly straight to within the errors in the measurements. There is a
suggestion of a bend in the trajectory of N5 after the 2006 epoch
(epoch 9 in Fig.~4), 
but this relies almost
entirely on the position at the 2008 epoch.  For
the more distant northern component (N2) there are kinks in the apparent
motion near epochs 1994, 2001 and 2006 (epochs 2, 7 and 9 in Fig.~4).  
There also appears to be an offset
in the 15 GHz positions (see Fig.~4).  This suggests a flatter spectrum on the
eastern side of this component.  A similar offset is suggested by the
multi-frequency trajectory plot of S2 based on the 5 and 8 GHz
observations, but since the motions are smaller and the component is
weaker it is difficult to be sure if the spectral gradient is 
real.  A kink is observed in component S2 near the 2003 epoch (epoch 8 
in Fig.~4).

In well studied core-jet sources, such as 3C\,345 (Zensus et al. 1995,
Unwin et al. 1997), jet components are found to travel along curved
trajectories and to change their speeds.  This acceleration has been
taken as justification for magnetically driven jets, as opposed to
purely hydrodynamical structures (Vlakhis \& Konigl 2004).  By analogy
with these core-jet sources, we draw a similar conclusion that the
jets of 1946+708 are likely to be magnetically driven.  Note, however,
that intrinsic velocities of jet components in core-jet AGNs are
typically in the range $\gamma \sim 3-10$, considerably larger than we
observe in the CSO 1946+708 of 1.3 $-$ 2.3 (see below).

Even before the core of 1946+708 was identified based on 15 GHz
observations, the first two epochs at 5 GHz suggested its location by
exhibiting bi-directional motions away from the center of the source.
Given these motions and the symmetry of the source (Figs.~1 and 2), it
is reasonable to assert that N5 and S5 were ejected at the same time,
and likewise N2 and S2.  Indeed, we are fortunate enough to witness
the ejection of N6 and S6 in $\sim$1997 as discussed above. 
Sensitive observations at 8.4 GHz (see Fig.~2), reveal that
there is continuous emission from the core out to both hot spots.  The
jet is probably not made up of discrete blobs that can be well
described by the elliptical Gaussian components that we identify and
modelfit.  Rather the jet appears to be a continuous flow, with
features of enhanced emission (shocks?)  that propagate down the jet.

Under the assumption that components are ejected in pairs, and that
all differences are due to Doppler boosting and light travel time
effects, we can use the observed
apparent velocity ratios and differences to solve for the intrinsic component
velocity and orientation.  
For simultaneously ejected components moving in opposite directions at
an angle $\theta$ to the line of sight at an intrinsic velocity $\beta$, it
follows directly from the light travel time difference that the ratio
of apparent projected distances from the origin ($d_{\rm a}$ for the
approaching side, $d_{\rm r}$ for the receding side) as well as the
ratio of apparent motions (approaching: $\mu_{\rm a}$, receding:
$\mu_{\rm r}$) is given at any time by (Taylor \& Vermeulen 1997):
\begin{equation}
{{\mu_{\rm a}}\over{\mu_{\rm r}}} = {{d_{\rm a}}\over{d_{\rm r}}} = 
\Biggr({{1+\beta\cos \theta}\over{1-\beta\cos \theta}}\Biggl)\,.
\end{equation}

Our other important constraint on the two parameters $\beta$ and
$\theta$ can be obtained from the separation rate $\mu_{\rm sep} =
|\mu_{\rm a}| + |\mu_{\rm r}|$, which, unlike $\mu_{\rm a} / \mu_{\rm
  r}$, is not subject to the uncertainty in the reference point. From
geometry and the conversion of angular to linear velocity we have:
\begin{equation}
v_{sep} = {\mu_{sep}\,D_a\,(1+z)} = {{2 \beta c \sin \theta}\over{(1 - \beta^2\cos^2 \theta)}}\,,
\end{equation}
where $v_{sep}$ is the projected separation velocity, $D_a$ is the
angular size distance to the source, and $z$ is the redshift.  Note
that Eq.~2 has a distance dependence, while Eq.~1 does not, so we can
solve the system jointly for the distance and hence the Hubble constant.

For the N5/S5 pair we find an apparent velocity ratio ${\mu_{\rm a}}/{\mu_{\rm r}}$
of 3.50 $\pm$ 0.44.  This leads to $\beta \cos{\theta}$ = 0.56 $\pm$ 0.04.  
At the same time, the apparent separation speed of the N5/S5 pair is 
1.40 $\pm$ 0.01 c.  
The two relations above are shown graphically in Fig.~6 (bottom panel).  
Assuming
a standard cosmology and H$_0$ = 71 km s$^{-1}$ Mpc$^{-1}$ we
find a common solution for an intrinsic velocity of 0.88 $\pm$ 0.03 c, at an inclination
of 50 $\pm$ 5\deg.  A similar analysis for the N2/S2 pair (top panel of Fig.~6), 
yields in intrinsic velocity of 0.57 $\pm$ 0.11 c  and with an inclination
of 36 $\pm$ 10\deg.  The intrinsic velocity therefore
changes from 0.9 c for N5/S5 to 0.6 c
for N2/S2 (a change of 2.7 $\sigma$).  There is no single intrinsic velocity that can fit the observations
for both pairs, and can be used to directly measure the Hubble constant.
We can, however, place a lower limit on the Hubble constant based
on the fact that the intrinsic velocity must be less than the speed
of light.  
We find that H$_0$ $>$ 57 km s$^{-1}$ Mpc$^{-1}$
in order to achieve a valid solution for the N5/S5 pair, 
and the weaker constraint that H$_0$ $>$ 28 km s$^{-1}$ Mpc$^{-1}$
in order to achieve a valid solution for the N2/S2 pair.
Since the velocities of N2/S2 differ from those of N5/S5 anyway, 
we see no reason why the latter should be particularly close 
to c, so for the rest of this discussion
we adopt H$_0$ = 71 km s$^{-1}$ Mpc$^{-1}$.

The solutions discussed above using equations 1 and 2 are based on
assuming that the components in a pair, as observed, are oppositely
directed and equally fast. We derive angles near 45\deg; the line of
sight depth difference between the components in a pair would therefore
be an appreciable fraction of their total distance along the connecting
line through the core, and roughly equal to the distance projected onto
the plane of the sky. We derive component speeds that are an appreciable
fraction of the speed of light. In combination, the speed and angle
solutions imply that the receding component, due to a significant light
travel time difference, should be observed at a rather younger age
(25\%-75\%) than its approaching counterpart, and at a correspondingly
smaller projected distance from the core. However, the observed arm
length ratios ($d_a/d_r$) are lower by a factor of about 2 compared to
the apparent velocity ratios (1.33 vs.\ 2.76 for N2/S2; and 1.80 vs.\
3.50 for N5/S5). This requires that currently the light travel time
differences between the approaching and receding components are less
significant with respect to the ages of the components than our simple
speed and angle solutions suggest. The arm length ratios depend on the
entire history of motion of the components, and the greater degree of
symmetry in arm lengths seems to imply that for a significant part of
their lifetime the components were slower and/or moving more along the
plane of the sky than they are now. A fully self-consistent
solution therefore will incorporate a model for the true core location
and for this evolution in velocity, since, inevitably, it implies that
the receding component is seen at a younger age and therefore with a
different velocity than its counterpart. This leads to an extension of
equations 1 and 2, into a set where the arm length ratio is directly
incorporated; helical jet models may provide a framework for this
extension. However, it is beyond the scope of this paper. To properly
verify such a model in the case of 1946+708 will require observations
spanning several decades. Nevertheless, we believe the velocities we
have derived, while not fully self-consistent, are indicative of the
fact that the jets of this source contain features that move with speeds
that are an appreciable fraction of the speed of light, and certainly
much faster than the advance speeds of the hot spots. Furthermore, the
differences in angle (by at least 15 degrees) as well as in speed (by at
least 50\%) between the N2/S2 and N5/S5 pairs are good indications of
the kinds of changes in component velocity that are evidently occurring
in and along the jets.  Both the discrepant arm length
ratios and the observed bending in projection of the overall jets on the
plane of the sky provide interesting constraints.


\subsection {Flux Density Evolution}

The time evolution of the flux densities of the N2/S2 and N5/S5
pair is shown in Fig.~7.  The N2 component has shown a steady
rise by $\sim$20\% over 13 years of monitoring at 8.4 GHz.
During that same time S2 has only slightly decreased.  
Meanwhile both N5 and S5 have been declining though
that appears to have leveled off starting in 2006.  The amount
of the decline was 75\% for N5 and 57\% for S5.  We are 
currently in the unbalanced situation where S5 appears to be 
brighter than N5, which would at first sight seem contrary
to the expectations from Doppler boosting.  However, we 
have to keep in mind that the light from S5 is delayed compared
to the light that we see from N5 
(by 27 years for the 2008 epoch according to the geometry derived 
above) so that the history of variability must
be taken into account when attempting to interpret flux
density ratios between component pairs.
A more detailed
analysis of the flux density ratios, taking into account the 
time variability, should eventually be possible.  Even so,
it may not be possible to explain the very significant 
changes in flux density shown in Fig.~7. given the stability 
of the component velocities (see Fig.~5).  Local circumstances
(e.g., variations in magnetic field strength or particle 
populations), and possible interactions of the jet components
with their environments, may influence the synchrotron emissivity
of the jet components. 

\subsection {Non-ballistic Motions}

The curvature in the jets of 1946+708 (Fig.~1) could be explained by
either (1) ballistic motion from a precessing nucleus; or (2) helical 
confinement of the components.  In the first case we would expect
components to travel in straight lines at constant apparent velocities.
In the latter case we would expect that components might 
travel along helical trajectories at apparent velocities that
changed in time.  Although there is no strong evidence yet
for deviations from constant apparent velocities, there do appear to
be some wiggles or kinks in the jets, and in particular if
one considers also the earliest 5 GHz measurement (C1 in 
Fig.~4), then there appears to have been a change in direction
of several degrees. This favors the second explanation of a 
helically confined jet.  Likewise, the presence of continuous
emission along the jet, well confined, even if traveling 
at different intrinsic velocities (N5 vs N2), supports the notional 
model of an intrinsically helical jet.


\subsection{The Advance Speed of the Hot Spots}

From the modelfit analysis it is difficult to ascertain the expansion
of the source.  Certainly it is clear that the motion is much slower
than the jet components.  Unfortunately the southern hot spot is
fairly faint and diffuse, so is correspondingly less amenable to
modelfitting than the compact jet components.  The northern hot spot
does have a bright and compact feature, and one can measure with
considerable accuracy its position relative to the core.  The observed
motion is $<$0.01 mas/y, at the level that one has to be
concerned about possible motions of the core component due to the
ejection of the N6/S6 pair.  Since the modelfit results are 
referenced to the core component, any apparent proper motion of the core 
translates into the addition of a systematic apparent velocity to 
all components.

An alternative approach is to compute a difference map between two
images, well separated in time.  Apparent motions then show up as
alternating positive and negative structures.  Fig.~8 shows such an
image constructed by differencing the images from the 1996 and 2008
epochs - both of which are of very high quality and low noise, and
were convolved with the same restoring beam.  Lighter regions in 
Fig.~8 indicate where the source was brighter in 2008.  In Fig.~8 one can see a
relatively clear signature of motion away from the core of the southern hot spot,
which appears as a positive (dark) structure (closer to the core) and a
negative (whiter) structure (further from the core) in the difference image.
The more rapidly moving jet components show similar structures.  From
measuring the advance of the edge of the southern hot spot we
estimate this apparent motion to be 0.008 $\pm$ 0.002 mas/y.  Since
the apparent motion measured is relative to the northern hot spot,
this is the overall apparent expansion rate of the source, and the
apparent motion of the southern hot spot would presumably be half this
value, or 0.004 mas/y, corresponding to a projected intrinsic velocity
of 0.024 c (7000 km/s).

Based on the measurement of the overall expansion rate, we can derive
a kinematic age for 1946+708 assuming a constant expansion rate.
The velocity measurement derived above yields a kinematic age
of 4000 $\pm$ 1000 years.  This is on the long end of CSO ages measured
to date (Gugliucci et al. 2005), which have been measured 
between 100 and 3000 years, though the statistics are admittedly
still poor.  Furthermore, there is a selection effect that 
slow expansions take longer to measure.  

The ratio of 15:1 for the northern to southern hot
spot flux densities (derived from modelfitting), is difficult
to explain in terms of Doppler boosting given the very low
apparent velocities measured for the hot spots. A more likely explanation
is that the interstellar medium
may be enhanced on the northern side, consistent with the observation
of greater HI opacity to the North (Peck \& Taylor 2001).  The arm
length ratio between northern and southern hot spots is close to 
parity, 0.94:1, with the northern hot spot being a little closer to the 
core. In the case of Doppler boosting we would expect the northern 
hot spot to be further away from the core, thus we favor a denser
medium on the northern side to be responsible for the difference
in flux densities and arm length ratios between the two hot spots.


\subsection{Depolarization by a Faraday Screen}

The low observed polarization (less than 0.4\% for the core, 
less than $\sim$0.3\% for the jet components, and less than 
0.1\% for the northern hot spot at 8.4 GHz in 1996), can be 
explained by Faraday depolarization due to ionized gas and a magnetic field
tangled on scales smaller than the angular resolution of
the observations.  This situation could naturally arise 
due to magnetic fields and free electrons associated with the
accretion disk.  Detection of polarization has only been found
in a few CSOs to date (Gugliucci et al. 2007).  Those few 
incidents of detected polarization occur in the approaching jets of CSOs
which are more asymmetric and core-dominated than typical 
CSOs, probably indicating a smaller angle to the line-of-sight,
and therefore a more shallow Faraday screen.

\section{Conclusions}

After a detailed, multi-frequency, multi-epoch study of the Compact
Symmetric Object 1946+708, we find the kinematic age of the extant
radio emission to be $\sim$4000 years.  On timescales of $\sim$10
years, outbursts occur producing symmetric components that emerge from
the core and travel at intrinsic speeds between 0.6 and 0.9 c towards
the hot spots.  Some of the individual jet components are observed to move 
along slightly bent, or kinked paths, but no components 
have yet been observed to change their speed.  The jet components
in general appear to be well confined, and to lie within the
overall ``S-symmetric'' shape of the jets.  We suggest that the jets are helically confined, rather
than ballistic in nature.  No linear polarization from the jets or
hot-spots is detected down to quite low levels (0.1 to 0.4\%),
consistent with observations of CSOs in general.

Future observations over the next decade should allow for a 
detailed analysis of the newly ejected component pair N6/S6.
Together with continued observations of N2/S2 and N5/S5 it should
be possible to measure accelerations in the jet components, and to tell
if each component pair has an intrinsic velocity that is 
established upon ejection.  A more sophisticated
analysis, taking into account the flux history of the jet components,
could provide a more stringent test of Doppler boosting in 
1946+708 and thereby ascertain the extent to which interactions
with the environment are important.  


\acknowledgments

We thank H. Smith for help with the data reduction of the 1998 and
1999 epochs.  G.B. Taylor gratefully acknowledges the University of
Bordeaux for hosting a visit during which much of this work was
undertaken.  This work has benefited from research funding from the
European Community's sixth Framework Programme under RadioNet R113CT
2003 5058187.  The National Radio Astronomy Observatory is operated by
Associated Universities, Inc., under cooperative agreement with the
National Science Foundation.  The European VLBI Network is a joint
facility of European, Chinese, South African and other radio astronomy
institutes funded by their national research councils.

\clearpage

\begin{figure}
\vspace{15cm}
\includegraphics{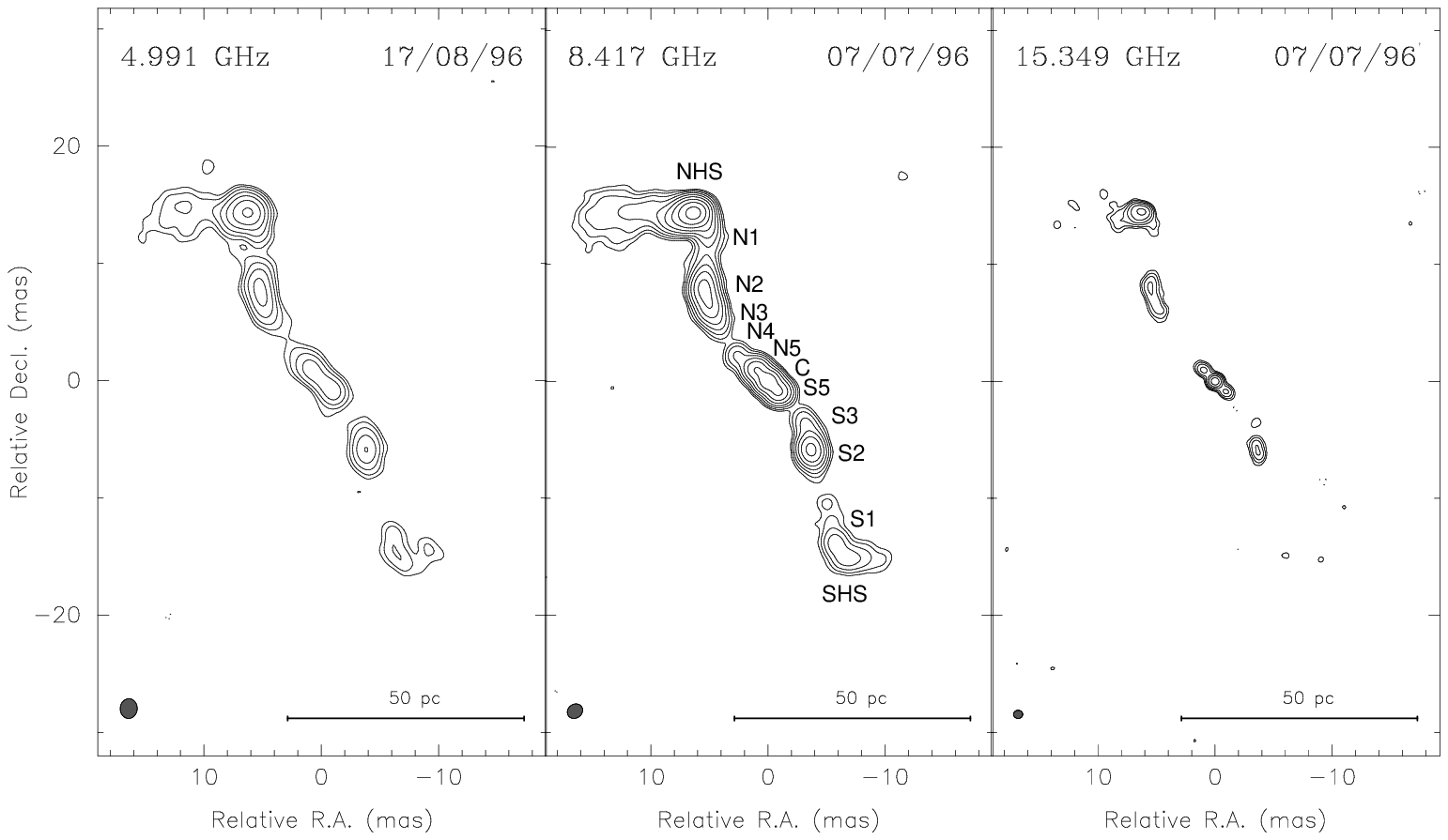}
\figcaption{Nearly contemporaneous VLBA observations of 1946+708 at 5, 
8 and 15 GHz.
The synthesized beam is drawn in the lower left-hand corner of
each plot and has 
dimensions: 1.68 $\times$ 1.45 mas in position angle $-$3.4\deg\ at 5
GHz; 1.37 $\times$ 1.18 mas in position angle $-$55\deg\ at 8
GHz; and 0.78 $\times$ 0.7 mas in position angle 90\deg\ at 15
GHz.
Contours are drawn logarithmically at factor 2 intervals with 
the first contour at 2, 0.25, and 0.75 mJy/beam at 5, 8 and 15 GHz
respectively.  The components labeled NHS and SHS in the 8 GHz image
are the northern
and southern hot spots respectively.  The component labeled ``C''
we identify as the core.  }
\end{figure}
\clearpage

\begin{figure}
\vspace{15cm}
\includegraphics{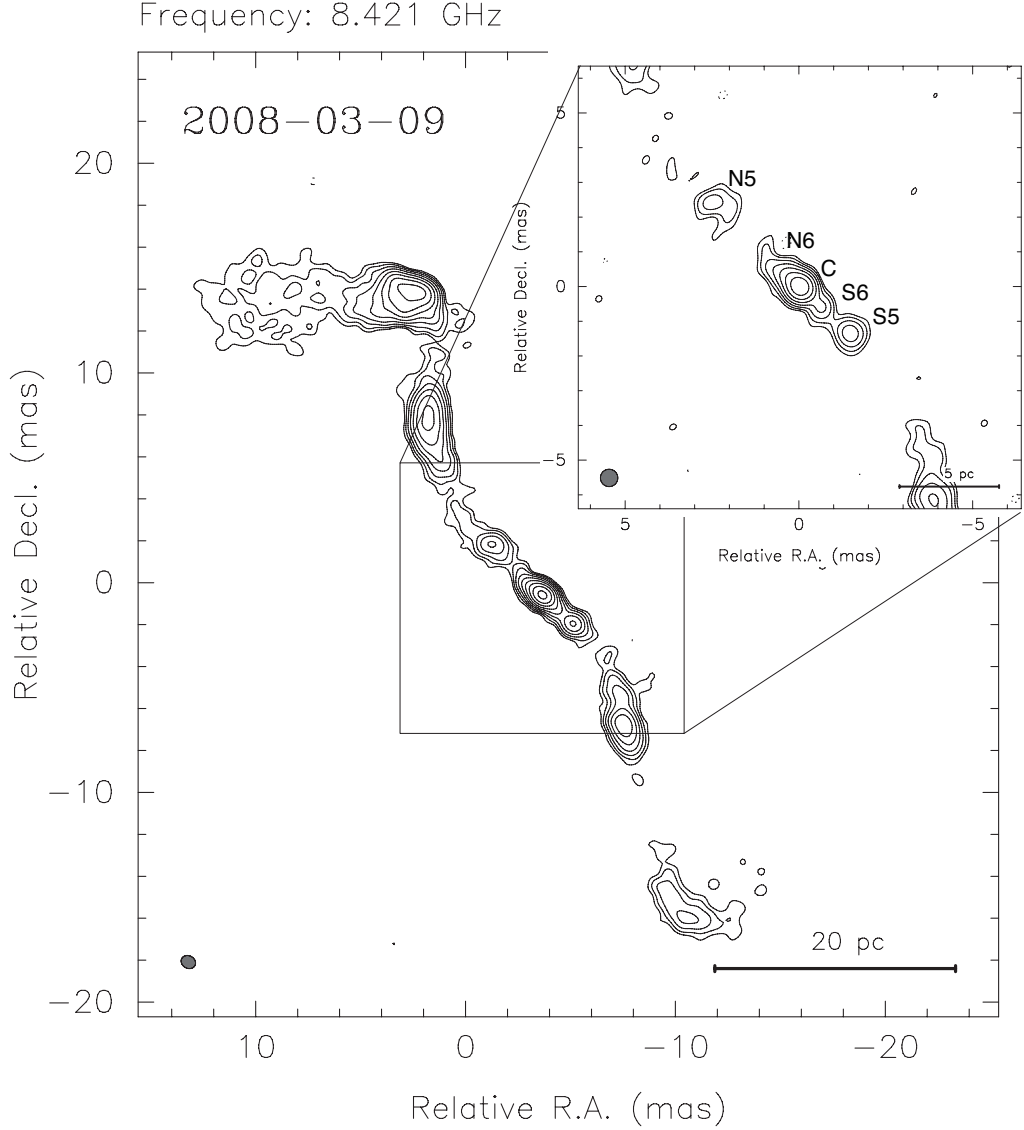}
\figcaption{The naturally weighted 8.4 GHz image from the 2008 epoch 
with a magnified view of the central region.  The central region has
full resolution and is uniformly weighted.
The synthesized beam is drawn in the lower left-hand corner of
each plot and has 
dimensions: 0.71 $\times$ 0.59 mas at position angle 62$^\circ$ for the entire image and 0.5 mas for the inset image.
Contours are drawn logarithmically at factor 2 intervals with 
the first contour at 0.2 mJy/beam for the entire image and at 0.5 mJy/beam for
the inset.  
A new pair of components is visible just north and just south
of the core. }
\end{figure}
\clearpage

\begin{figure}
\vspace{15cm}
\includegraphics{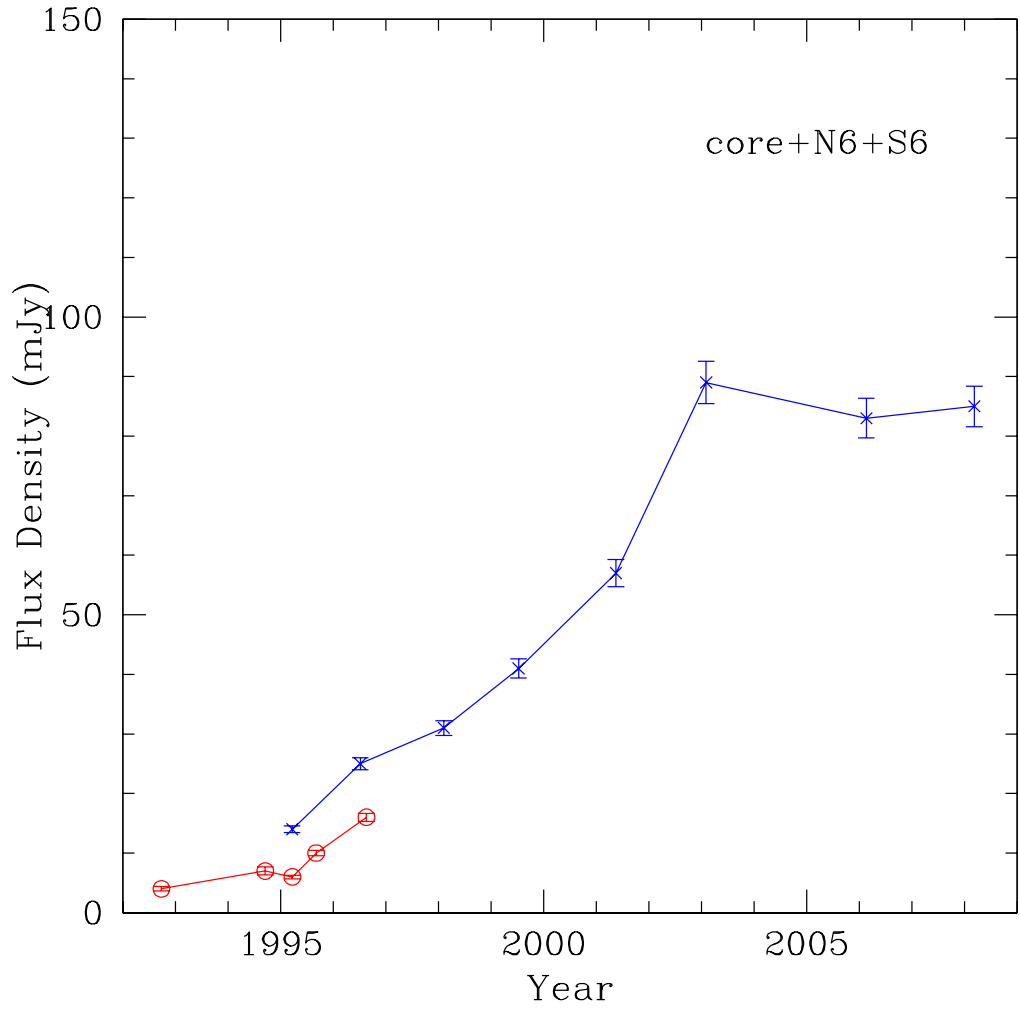}
\figcaption{The flux densities of the core component 
between 1992 and 2008 as derived from Gaussian modelfitting.
The discontinuity is the result of the first five monitoring epochs 
being at 5 GHz (red), while the last eight epochs have used 8.4 GHz (blue).
}
\end{figure}
\clearpage

\begin{figure}
\vspace{15cm}
\includegraphics{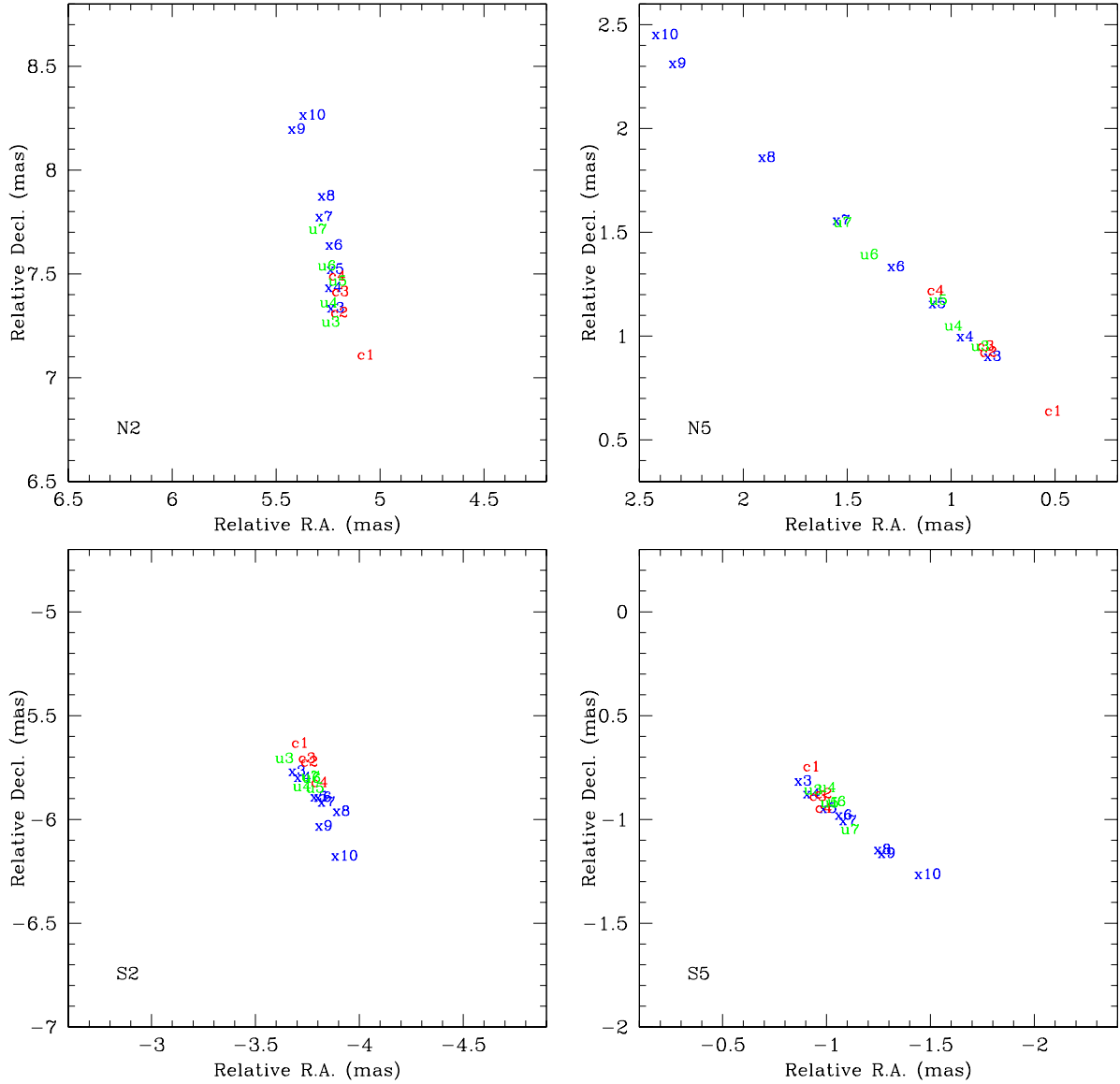}
\figcaption{The path of the 4 paired jet components (N2/S2 and N5/S5)
between 1992 and 2008 as derived from Gaussian modelfitting.  Letters and
colors designate the frequency band with red ``c'' for 5 GHz, blue``x'' 
for 8.4 GHz, and green ``u'' for 15 GHz.
The numbers correspond to the position of the component at epochs 1 = 1992
Sep; 2 = 1994 Sep; 3 = 1995 Mar or Sep; 4 = 1996 Jul or Aug; 
5 = 1998 Feb; 6 = 1999 Jul; 7 = 2001 May; 8 = 2003 Feb; 9 = 2006 Feb; 10 = 2008 Mar. }
\end{figure}
\clearpage

\begin{figure}
\vspace{15cm}
\includegraphics{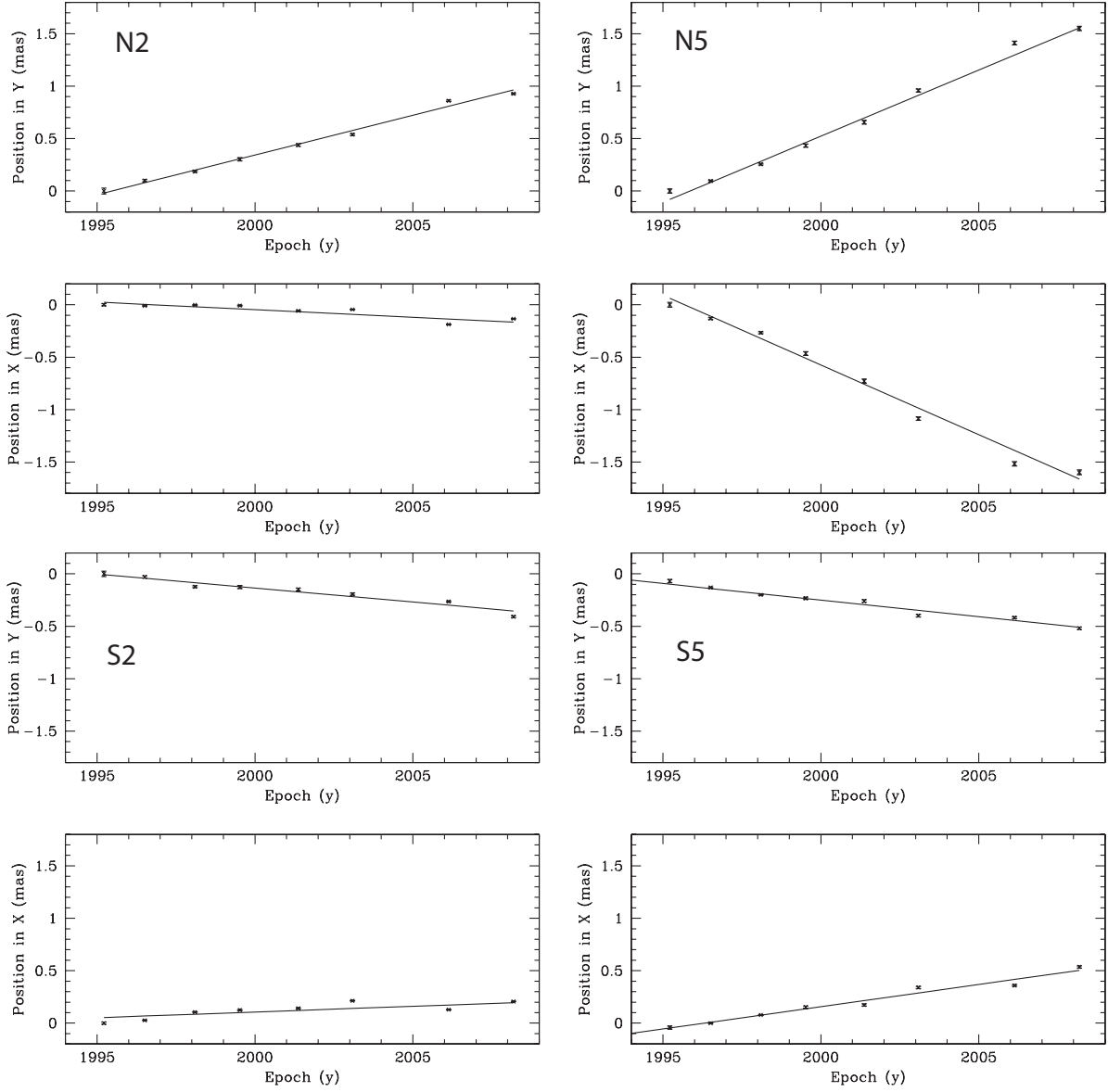}
\figcaption{Velocities of N2, S2, N5 and S5 derived by
fitting to the X and Y components of motion observed at 8.4 GHz.  Velocities are tabulated
in table 2.
The zero point is taken to be the position of the component at
the first epoch. }
\end{figure}
\clearpage

\clearpage
\begin{figure}
\vspace{19cm}
\includegraphics{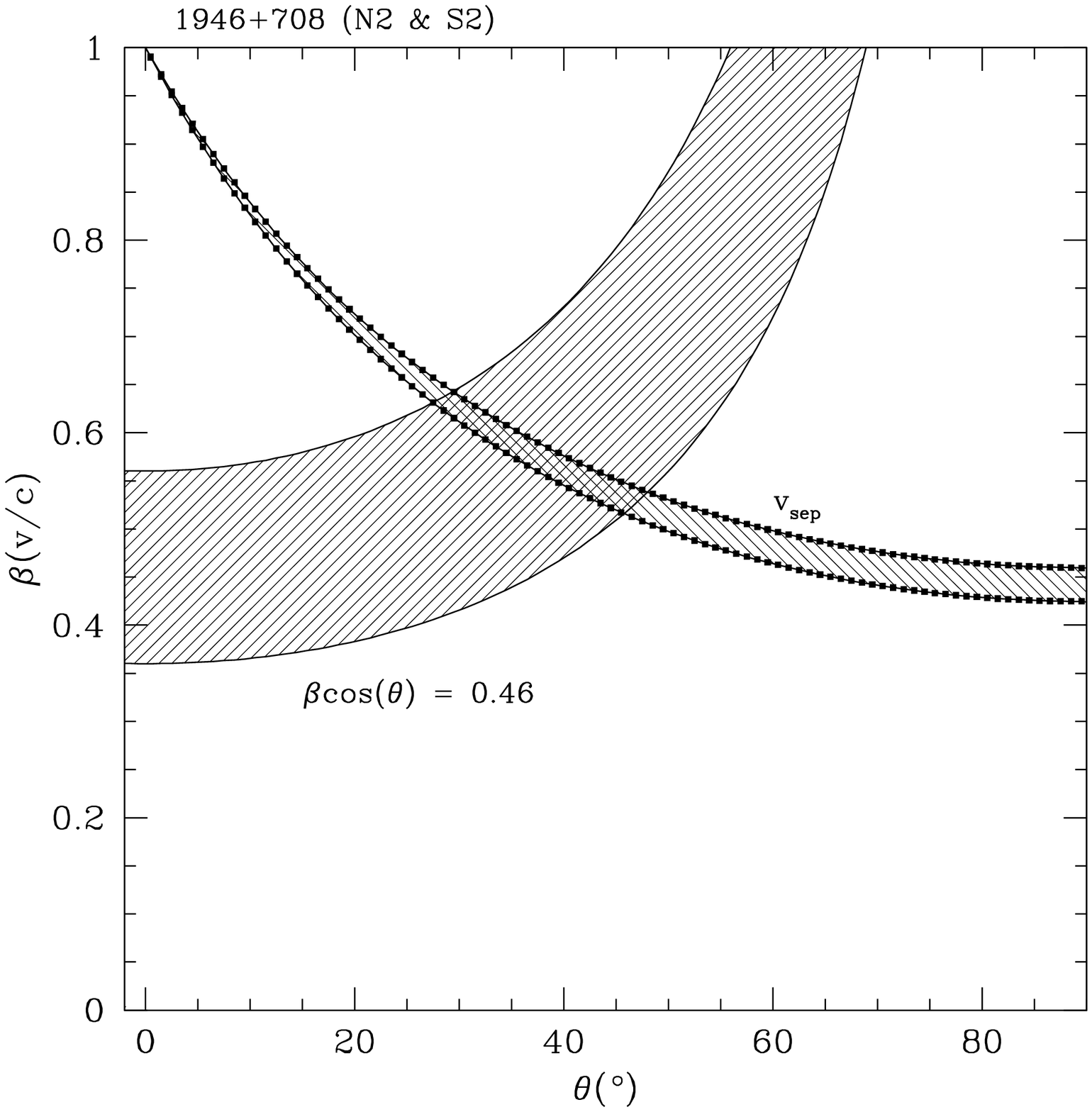}
\includegraphics{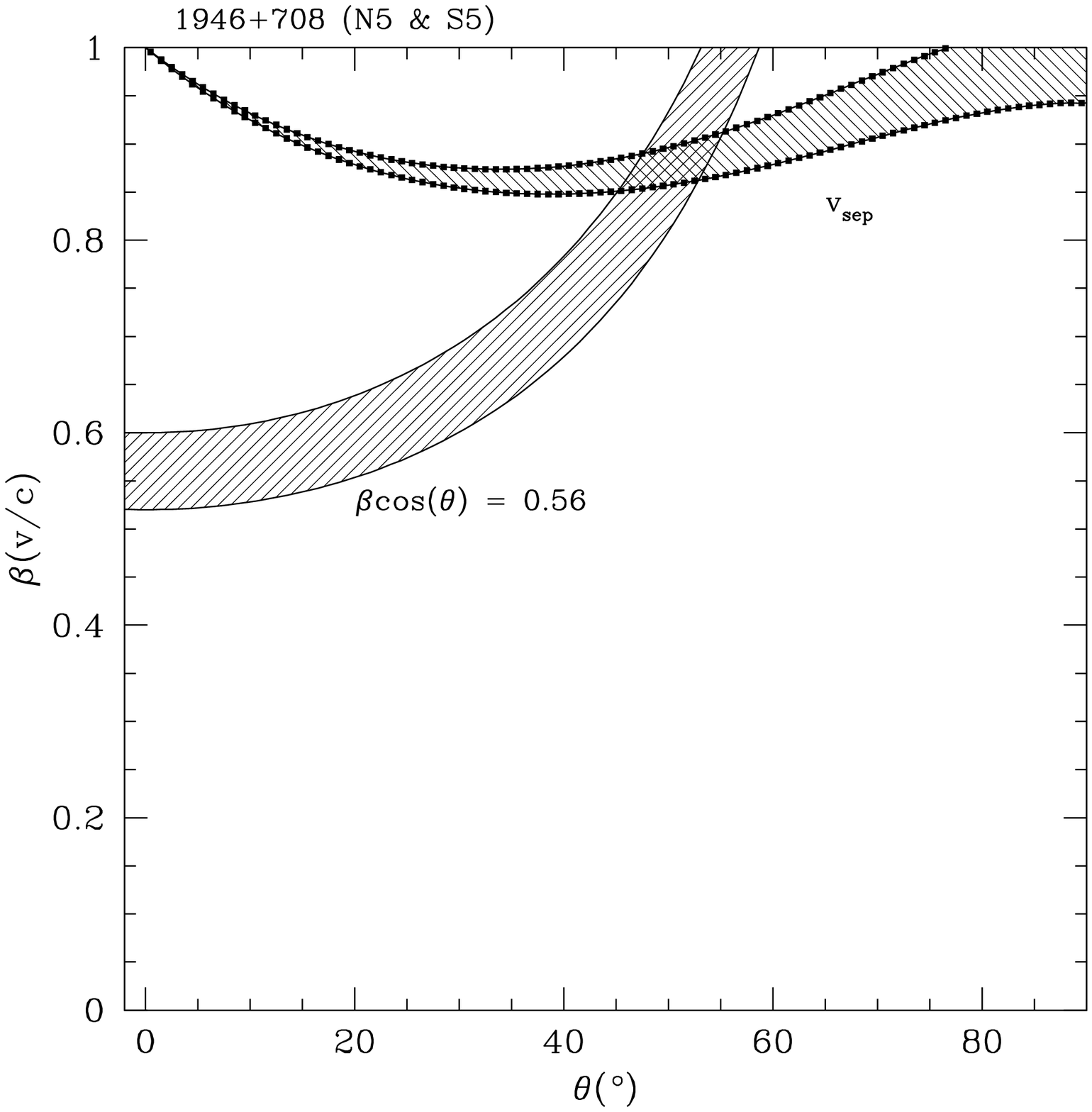}
\figcaption{The intrinsic jet velocity ($\beta$) plotted against the inclination
of the source ($\theta$) measured from the line-of-sight to the jet
axis.  The solid lines represent the constraint $\beta \cos\theta =
0.46 \pm 0.1$ from the apparent velocity ratio $\mu_a/\mu_r = 2.76 \pm 0.69$
of components N2 and S2 (top). The heavy dashed lines show the constraint from the apparent separation velocity $v_{sep} = 0.63 \pm 0.01$ c for N2 and S2 with 
$H_0$ = 71 km s$^{-1}$ Mpc$^{-1}$ assumed. 
At bottom is the same plot derived from apparent velocity measurements 
of the N5 and S5 pair ($\mu_a/\mu_r = 3.50 \pm 0.44$, 
$\beta\cos\theta = 0.56 \pm 0.04$ and $v_{sep} = 1.40 \pm 0.01$ c)
}
\end{figure}

\begin{figure}
\vspace{15cm}
\includegraphics{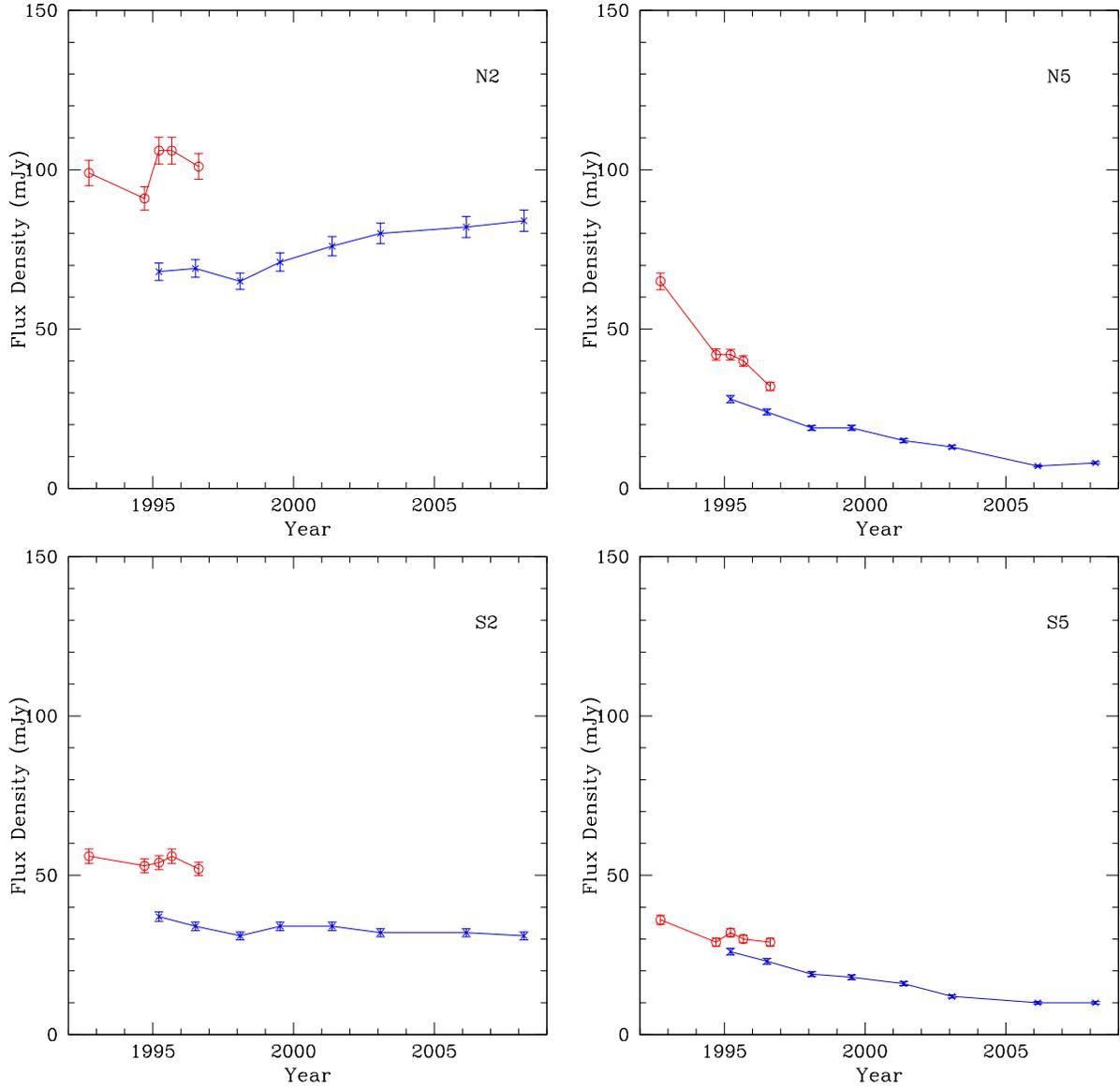}
\figcaption{The flux densities at 5 GHz (red), and 8.4 GHz (blue), of the 
paired jet components N2/S2 and N5/S5
between 1992 and 2008 as derived from Gaussian modelfitting.}
\end{figure}
\clearpage

\begin{figure}
\vspace{15cm}
\includegraphics{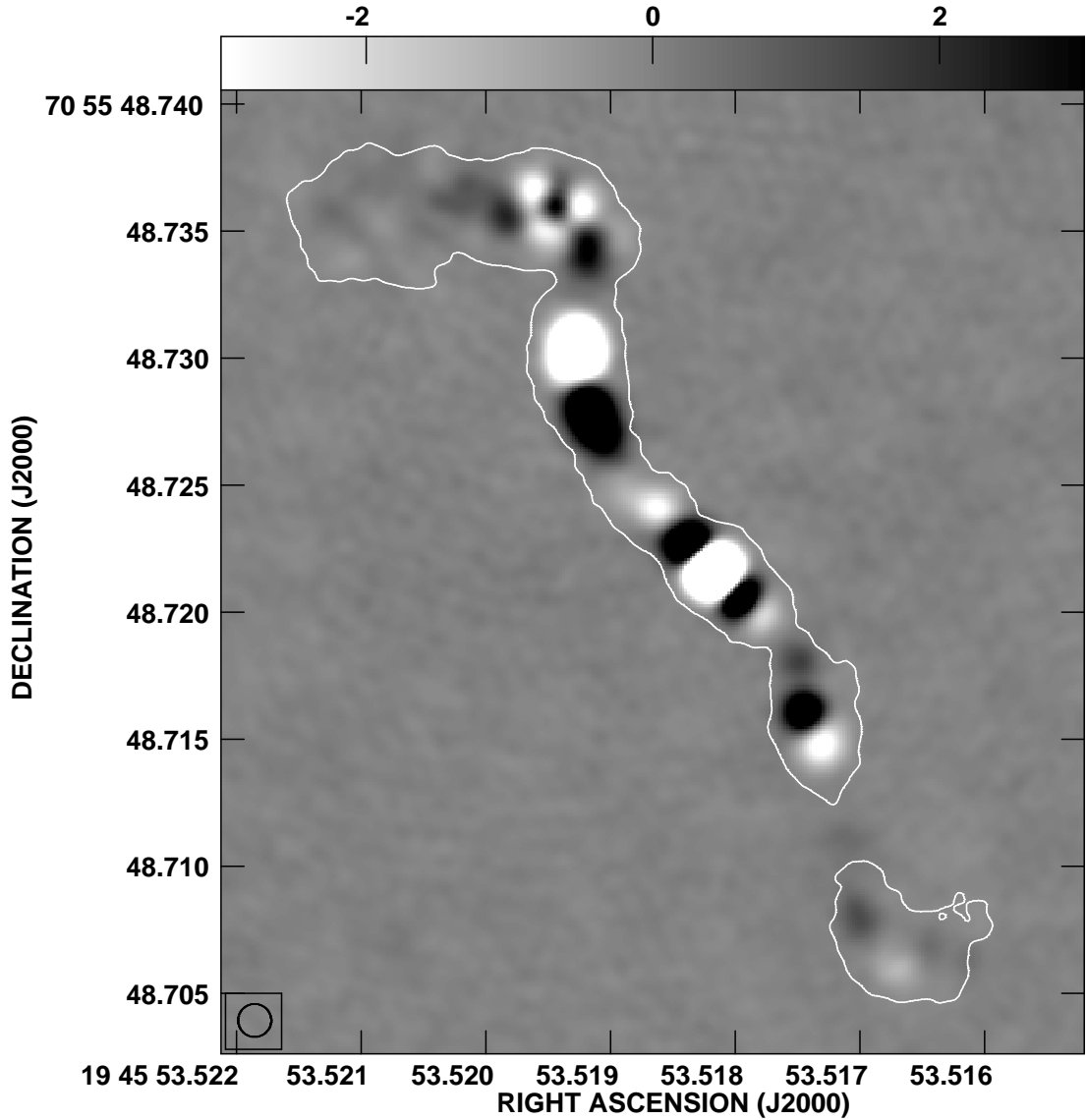}
\figcaption{Difference image between the 1996 and 2008 epochs at 8.4 GHz.
Both images were restored using a 1.3 mas circular beam,
and the images were registered to the position of the northern hot 
spot using a two-dimensional gaussian fit implemented in the 
AIPS task JMFIT.  The grey scale range is from $-$3 to 3 mJy/beam, and the 
single contour level is from the 2008 epoch at 0.25 mJy/beam. 
Lighter regions in
indicate where the source was brighter in 2008. 
}
\end{figure}
\clearpage

\def\dg{$^{\circ}$}
\begin{center}
Table 1.  O{\sc bservational} P{\sc arameters for} 1946+708
\scriptsize
\vskip2pt
\begin{tabular}{r r r r r r r r r}
\hline
\hline
Date & Frequency & Bandwidth & No.Ant & Scan Length & Total Time & rms\\
 & (GHz) & (MHz) & & (min) & (hours) & ($\mu$Jy/b) \\
\hline
\noalign{\vskip2pt}
1992 Sep 24 & 5.0 & 2 & 15 & 20 & 0.92 & 340\\
1994 Sep 15 & 5.0 & 2 & 12 & 20 & 0.95 & 588\\
1995 Mar 22 & 5.0 & 16 & 10 & 8 & 0.93 & 177\\
            & 8.4 & 16 & 10 & 2 & 3.83 & 163\\
            & 15.4 & 16 & 10 & 8 & 2.13 & 315\\
1995 Sep 3 & 5.0 & 8 & 10 & 6.5 & 0.98 & 235\\
1996 Jul 7 & 8.4 & 32 & 9 & 2 & 5.48 & 57 \\
           & 15.4 & 32 & 9 & 12 & 1.34 & 218\\
1996 Aug 18 & 5.0 & 8 & 9 & 5.5 & 0.73 & 251\\
1998 Feb 6 & 8.4 & 32 & 10 & 2 &  4.67 & 45 \\
           & 15.4 & 32 & 10 & 10 & 1.21 & 99\\
1999 Jul 11 & 8.4 & 32 & 9 & 2 & 2.74 & 87 \\
            & 15.4 & 32 & 8 & 10 & 0.78 & 161 \\
2001 May 17 & 8.4 & 32 & 10 & 2 & 3.49 & 76\\
           & 15.4 & 32 & 10 & 10 & 1.74 & 250\\
2003 Feb 2$^1$ & 8.4 & 32 & 14 & 4 & 6.67 & 55 \\
2006 Feb 18 & 8.4 & 32 & 14 & 4 & 3.33 & 38 \\
2008 Mar 9 & 8.4 & 32 & 14 & 4 & 3.33 & 48 \\
\hline
\end{tabular}\\
\end{center}
\vskip-12pt
$^1$ 2003 Feb 2 is the mean epoch for observations taken with
the VLBA alone on January 24, 2003 and with a Global array 
consisting of the VLBA, Westerbork phased array, 
Onsala, Medicina and Noto on 2003 February 10.  The observations
were combined for calibration and imaging.
\smallskip
\addtocounter{table}{1}
\begin{center}
\begin{deluxetable}{lccccccc}
\tabletypesize{\scriptsize}
\tablecolumns{8}
\tablewidth{0pt}
\tablecaption{Component Relative\tablenotemark{1}\ ~Motion Fitting Results\label{Motions_Table}}
\tablehead{\colhead{Component}&\colhead{Velocity in $y$}&\colhead{$\chi^{2}$}&\colhead{Velocity in $x$}&\colhead{$\chi^{2}$}&\colhead{Velocity}&\colhead{Velocity}&\colhead{Angle of Motion}\tablenotemark{2} \\
\colhead{} & \colhead{(mas/y)} & \colhead{}&\colhead{(mas/y)}&\colhead{}&\colhead{(mas/y)}&\colhead{(c)}&\colhead{($^o$)}}
\startdata
core & Reference Component &... &... &... &... &... &...  \\
N2 & \phs0.076 $\pm$ 0.0008 & 0.90 & $-$0.014 $\pm$ 0.0002 & 3.0 & 0.077 $\pm$ 0.0008 & 0.461 $\pm$ 0.005 & 10.4  \\
S2 & $-$0.026 $\pm$ 0.0008 & 0.80 & \phs0.011 $\pm$ 0.0004 & 3.0 & 0.028 $\pm$ 0.0009 & 0.167 $\pm$ 0.005 & $-$157.1 \\
N5 & \phs0.126 $\pm$ 0.0013 & 1.4 & $-$0.132 $\pm$ 0.0014 & 2.3 &  0.182 $\pm$ 0.0019 & 1.088 $\pm$ 0.011 & 46.3  \\
S5 & $-$0.031 $\pm$ 0.0008 & 0.53 & \phs0.042 $\pm$ 0.0007 & 2.1 & 0.052 $\pm$ 0.001 & 0.311 $\pm$ 0.006  & $-$126.4  \\
\enddata
\vskip-5pt
\tablenotetext{1}{Errors quoted are for apparent motions relative to the core only, and do not 
include any motion of the core component.  The systematic error that would
be produced by motion of the core is 
estimated by the slow apparent relative motions of the northern hot spot (NHS)
at $<$0.005 mas/yr in both Right Ascension and Declination.
Only the 8.4 GHz modelfit 
results are used.   }
\tablenotetext{2}{Angles measured from north through east.}
\end{deluxetable}
\end{center}

\clearpage

\begin{references}

\reference{giu05} Gugliucci, N.E., Taylor, G.B., Peck, A.B., \& Giroletti, M. 2005 ApJ, 622, 136
\reference{giu07} Gugliucci, N.E., Taylor, G.B., Peck, A.B., \& Giroletti, M. 2007 ApJ, 661, 78
\reference{bob81} Hjellming, R.~M., \& Johnston, K.~J.\ 1981, \apjl, 246, L141 
\reference{fom99} Fomalont, E.~B.\ 1999, Synthesis Imaging in Radio Astronomy II, 180, 301 
\reference{ows98} Owsianik, I. \& Conway, J.~E. 1998, A\&A, 337, 69
\reference{pec00} Peck, A.~B., Taylor, G.~B., Fassnacht, C.~.D., 
Readhead, A.~C.~S., \& Vermeulen, R.~C. 2000, ApJ, 534, 104
\reference{pec01} Peck, A.B., \& Taylor, G.B. 2001, ApJL, 554, L147
\reference{pil03} Pihlstr\"om, Y.~M., Conway, J.~E., \& Vermeulen, R.~C. 2003, A\&A, 404, 871
\reference{rea96} Readhead, A.~C.~S., Taylor, G.~B., Xu, W., Pearson, T.~J.,
Wilkinson, P.~N., \& Polatidis, A.~G. 1996, ApJ, 460, 612
\reference{rob08} Roberts, D.~H., Wardle, J.~F.~C., Lipnick, S.~L., Selesnick, P.~L., 
\& Slutsky, S.\ 2008, \apj, 676, 584 
\reference{sch83} Schwab, F.~R., \& Cotton, W.~D. 1983, AJ, 88, 688
\reference{she94} Shepherd, M.~C., Pearson, T.~J., \& Taylor, G.~B. 1994, BAAS, 26, 987
\reference{she95} Shepherd, M.~C., Pearson, T.~J., \& Taylor, G.~B. 1995, BAAS, 27, 903
\reference{tay94} Taylor, G.~B., Vermeulen, R.~C., Pearson, T.~J., Readhead, A.~C.~S., Henstock, D.~R., Browne, I.~W.~A., \& Wilkinson, P.~N. 1994, ApJS, 95, 345
\reference{tay96} Taylor, G.~B., Readhead, A.~C.~S., \& Pearson, T.~J. 1996, ApJ, 463, 95 
\reference{tay97} Taylor, G.~B., \& Vermeulen, R.~C. 1997, ApJL, 485, L9
\reference{tay00} Taylor, G.~B., Marr, J.~M., Readhead, A.~C.~S., \& Pearson, T.~J. 2000, ApJ, 541, 112 
\reference{unw92} Unwin, S.~C. Wehrle, A.~E., Lobanov, A.~P., Zensus, J.~A., 
Madejski, G.~M., Aller, M.~F., \& Aller, H.~D. 1997, ApJ, 480, 596
\reference{Vla04} Vlahakis, N. \& Konigl, A. 2004, ApJ, 605, 656
\reference{Zen95} Zensus, J.~A., Cohen, M.~H., \& Unwin, S.~C. 1995, ApJ, 443, 35

\end{references}
\end{document}